\documentclass[letterpaper]{article} 
\usepackage{aaai2026}  
\usepackage{times}  
\usepackage{helvet}  
\usepackage{courier}  
\usepackage[hyphens]{url}  
\usepackage{graphicx} 
\usepackage{natbib}  
\usepackage{caption} 
\usepackage{algorithm}
\usepackage{algorithmic}
\usepackage{newfloat}
\usepackage{listings}
\DeclareCaptionStyle{ruled}{labelfont=normalfont,labelsep=colon,strut=off} 
\floatstyle{ruled}
\newfloat{listing}{tb}{lst}{}
\floatname{listing}{Listing}
\title{Multimodal Data Fusion to Capture Dynamic Interactions between Built Environment and Vulnerable Older Adults}
\author {
    Houhao Liang\textsuperscript{\rm 1},
    Azrin Jamaluddin\textsuperscript{\rm 2},
    Kresimir Friganovic\textsuperscript{\rm 2},
    Kirstie Neo\textsuperscript{\rm 2},
    Raphael Han\textsuperscript{\rm 2},
    Navrag Singh\textsuperscript{\rm 2},
    Panos Mavros\textsuperscript{\rm 3}
}
\affiliations {
    \textsuperscript{\rm 1}CNRS@CREATE, 1 CREATE Way, CREATE Tower, Singapore\\
    \textsuperscript{\rm 2}Singapore-ETH Center, 1 CREATE Way, CREATE Tower, Singapore\\
    \textsuperscript{\rm 3}I3, UMR-9217 CNRS, Télécom Paris, Palaiseau, France\\
    houhao.liang@cnrsatcreate.sg, azrin.jamaluddin@sec.ethz.ch, kresimir.friganovic@sec.ethz.ch, kirstie.neo@sec.ethz.ch, raphael.han@sec.ethz.ch, navrag.singh@sec.ethz.ch, panos.mavros@telecom-paris.fr
}

\usepackage{bibentry}

\begin{document}

\maketitle

\begin{abstract}
Ensuring safe and inclusive mobility for vulnerable older adults is an emerging priority in urban planning. However, existing data sources such as surveys or GIS-based audits provide limited insight into how micro-scale built environment (BE) features influence real-world behavior and perception. This study presents a novel multimodal data-fusion approach that integrates wearable and environmental sensing to dynamically represent human-environment interactions and quantify the BE impacts on mobility among vulnerable older adults, specifically those with knee osteoarthritis or a history of falls. Data collected during naturalistic walking sessions in Singapore, are used to demonstrate this framework of synchronized streams from eye tracking, kinematic sensors, physiological monitors, GPS, and video recordings. Preliminary results show how AI-driven data fusion can uncover behaviorally and perceptually significant urban segments, providing a basis for actionable insights in inclusive design. This human-centered analytical approach advances the representation of urban environments from the perspective of vulnerable pedestrians, establishing a foundation for evidence-based, age-friendly city planning.

\end{abstract}


\section{Introduction}
As cities evolve toward human-centered and data-driven design, urban planners increasingly recognize the importance of understanding how built environment (BE) features shape everyday human experiences \cite{oecd:1}. For vulnerable populations, particularly older adults with mobility impairments, such as those living with knee osteoarthritis (OA), or a history of falls, the BE can determine whether a space feels walkable, supportive, or hazardous. Designing inclusive and accessible BE is therefore essential to sustain mobility and promote social engagement among older adults. 

Achieving this vision requires accurate, data-driven representations of how individuals with mobility challenges dynamically interact with their surroundings in real-world contexts. Urban planning increasingly depends on the representation and quantification of urban environments to support evidence-based design and policy decisions. However, data used in current planning practices are often derived from surveys, interviews, and field audits, which provide subjective and aggregated perspectives. Such methods, while informative at the neighborhood scale, cannot capture the moment-to-moment behavioral, perceptual and physiological responses that occur during real-world walking. As a result, they provide limited understanding of how specific BE features influence physical adaptation, and walking stability, particularly among vulnerable older adults with mobility impairments.

Motivated by recent advances in wearable sensing, computer vision, and robotics, new opportunities have emerged to represent and quantify urban experience through multimodal data fusion. Integrating visual, gaze, gait, and physiological signals, we examine how the BE dynamically influences human movement and perception at high temporal and spatial resolutions. Such evidence-based metrics can assist planners in evaluating comfort, safety, and accessibility at the micro scale, thereby complementing traditional environmental data. In this context, the present study introduces a multimodal data-fusion approach to dynamically represent and quantify BE and investigate the impacts on mobility and perception among vulnerable older adults. Through the combination of wearable and environmental sensing in real urban settings, the study provides an moment-to-moment representation of human–environment interaction that extends beyond static urban models.

\begin{figure*}[t]
\centering
\includegraphics[width=1\textwidth]{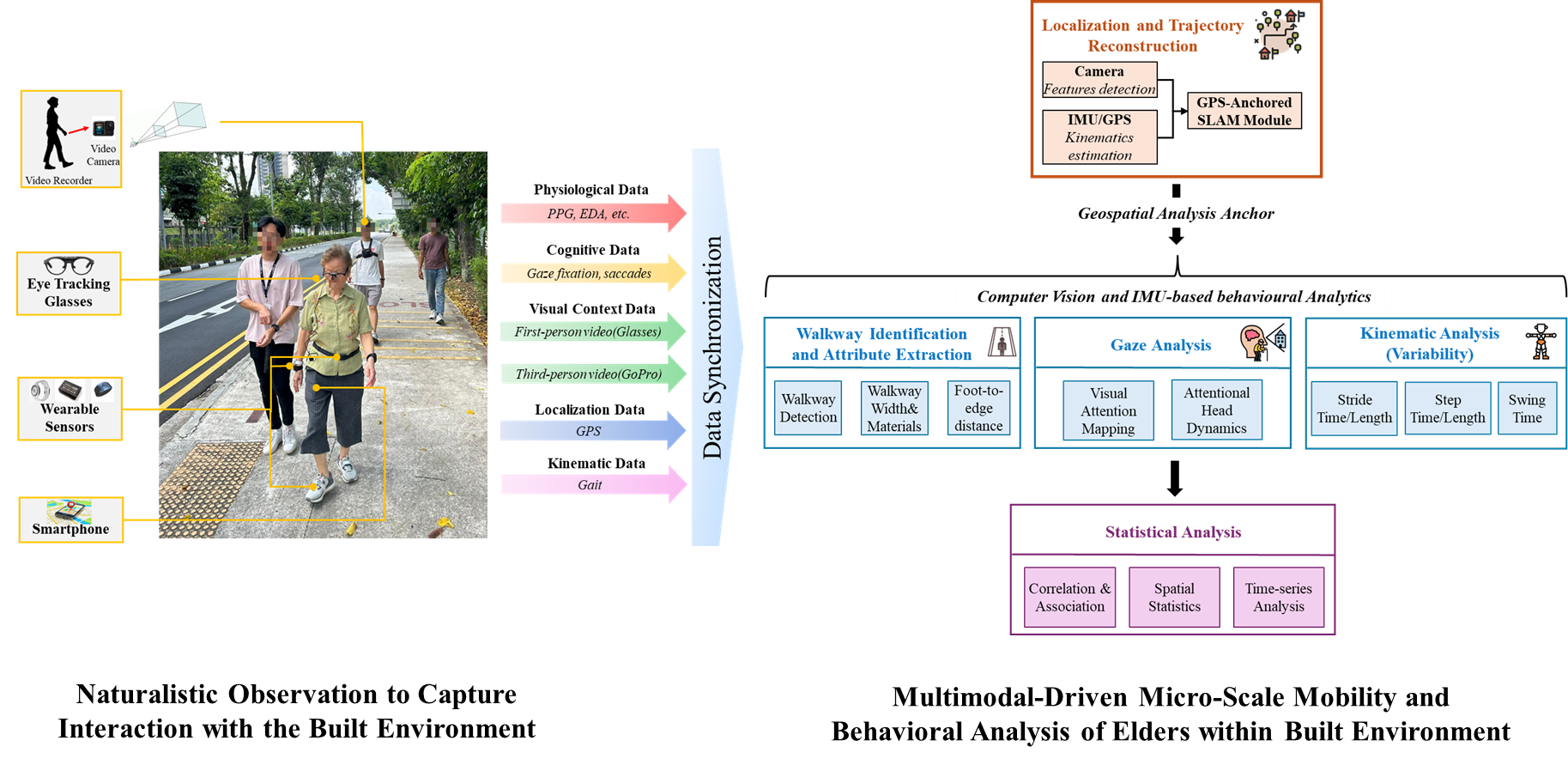} 
\caption{Overview of the multimodal data collection setup used to represent built-environment impacts on older adults’ mobility. Participants were equipped with eye-tracking glasses, wearable physiological sensors, and body-mounted IMUs while walking along urban routes in Singapore. A trailing GoPro camera provided third-person video. The integrated configuration captures synchronized gaze, gait, physiological, and environmental data for dynamic and fine-grained analysis of human–environment interactions.}
\label{fig1}
\end{figure*}

\section{Related Work}
As populations age, older adults often experience declines in cognitive function and perceptual acuity that influence how they interact with their surrounding environment \cite{RN19}. In this context, understanding and supporting these interactions have become growing priorities in both urban planning and public health \cite{RN15}. Built environment (BE) features, including pavement conditions, pedestrian crossings, lighting, streetscape elements, and spatial layout, can substantially affect mobility behavior among older adults, either facilitating or constraining safe and confident movement in daily life \cite{RN16, RN17}. Given these impacts, it is essential to capture how BE features shape older adults’ mobility and behavior. 

Conventional data sources to support urban design such as interviews, surveys, and field audits have provided valuable insights into walkability and accessibility \cite{RN22,RN21}. However, these methods depend on self-reported experiences, which can be affected by recall bias, declining response rates, and the substantial time demands placed on participants and researchers. Spatial analysis techniques based on geographic data represent another category of assessment. These geospatial approaches utilize diverse datasets, such as environmental sensing (e.g. traffic noise) \cite{RN27}, GPS logs \cite{RN25}, and Street View Imagery \cite{RN26}, to map and model BE features across large urban areas. Yet, these methods often rely on large-scale aggregated statistical approaches, and thus inadequately capture person-environment interactions at the micro-level, or account for individual differences. Hence, such assessments may overlook discrepancies in how environmental conditions affect different individuals \cite{RN17}.

More recently, individual-centered sensing has emerged as a promising method to capture how people interact with the BE in real-world settings, with the additional advantage of scalability for large population studies \cite{RN29}. Participants can be equipped with wearable devices that continuously record real-time physiological (e.g., heart-rate variability (HRV), electrodermal activity (EDA)), kinematic (e.g., gait dynamics), and cognitive (e.g., gaze behavior) data. For example, motion sensors placed on the feet or waist have been used to quantify gait patterns and detect subtle changes in walking behavior in response to environmental challenges such as uneven pavement or slope \cite{RN39, RN10}. Similarly, wrist-worn sensors measuring HRV, EDA, and skin temperature effectively capture physiological stress responses to environmental stressors, such as crowded crossings or architectural obstacles \cite{RN28}. 

Despite the increasing adoption of data-driven methods to investigate the BE’s impact, a critical gap remains in the integration of such multimodal data for micro-scale and holistic analysis. Few studies have comprehensively combined visual, gaze, gait, and spatial information to understand how older adults with mobility impairments interact with detailed BE features during everyday walking. Addressing this gap, the present study aims to develop a data-driven representation of BE features and quantify their micro-scale impacts on mobility and behavior among vulnerable older adults. Leveraging advances in computer vision, robotics, and spatial mapping, this study enables dynamic and fine-grained representation of BE features in real urban contexts.

\section{Study Overview}
The study focuses on community-dwelling older adults aged 45 to 90 years living in the Yio Chu Kang district of Singapore. This area represents a dense residential environment typical of mature Singapore neighborhoods. The selected walking routes include covered walkways, pedestrian crossings, green corridors, and commercial access paths, providing a realistic testbed for examining how built environment (BE) features influence movement and perception in older adults. Participants were purposively recruited through healthcare institutions and community networks. Eligibility required either a clinical diagnosis of knee osteoarthritis (OA) or at least one fall within the past two years, both of which are associated with reduced balance confidence and altered gait. The participants thus represent a vulnerable population segment that frequently encounters physical and cognitive challenges during daily mobility.

A naturalistic observation protocol is designed to capture real-world physical, perceptual, and cognitive interactions with the BE using high-resolution multimodal data, including gaze, video, motion, and physiological signals. An example of the multimodal setup is shown in Figure \ref{fig1}. Each participant walks along a self-selected outdoor walking route while being observed, with sessions typically lasting 60 to 90 minutes. ZurichMOVE inertial measurement units (IMUs) are attached to the feet, wrists, trunk, and head to enable full-body kinematic analysis \cite{RN10}. Additionally, participants wear an Axivity A6 \cite{axivity} on the lower back to record high-frequency accelerometer and gyroscope data, supporting the analysis of gait parameters such as step count, variability, and symmetry. Participants also wear Pupil Labs Neon eye-tracking glasses \cite{RN1}, which records gaze vectors, head kinematics, and first-person video to support in-depth analysis of visual attention and interaction with BE features such as signage, pathways, and barriers. Physiological signals are captured using the Empatica EmbracePlus wristband, which records EDA, skin temperature, and heart rate, laying the foundation to obtain insight into stress and autonomic regulation during walking. 

To complement the first-person perspective, a researcher follows each participant at a distance, recording third-person video using a GoPro camera. This footage provides external context, capturing how participants move through and orient themselves within the environment. It also serves as a secondary source for environmental annotation and behavioral interpretation. All data streams are time-synchronized, enabling integrated multimodal analysis of spatial trajectories, gaze patterns, gait characteristics, and older adults environment interactions at a high temporal resolution.

\begin{figure}[t]
\centering
\includegraphics[width=1\columnwidth]{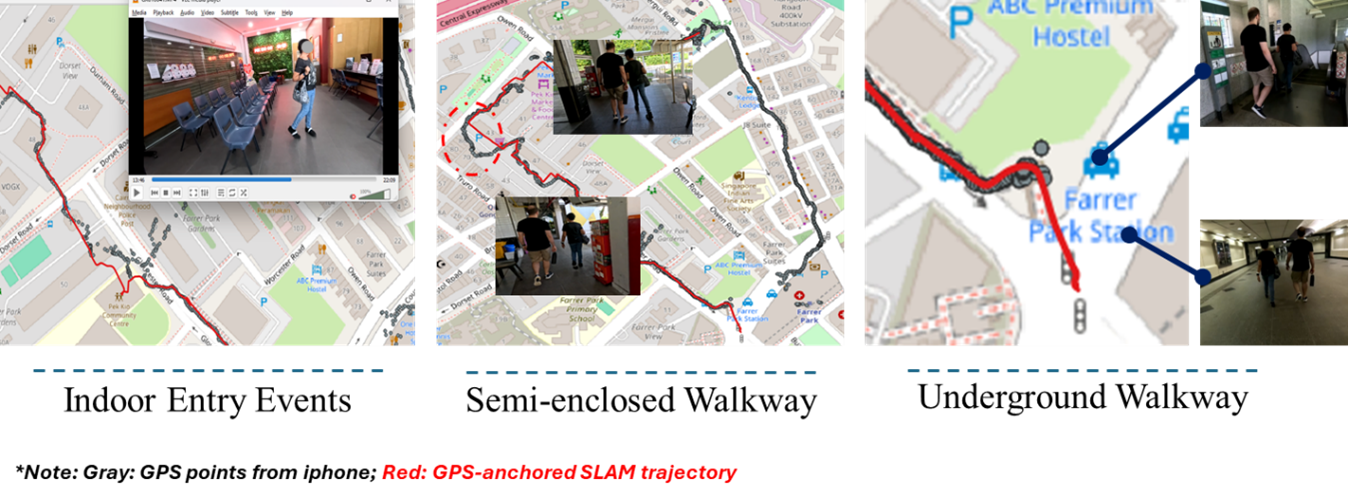} 
\caption{Activities in GPS denied environment while SLAM trajectory with GPS data points.}
\label{fig2}
\end{figure}

\begin{figure*}[t]
\centering
\includegraphics[width=1\textwidth]{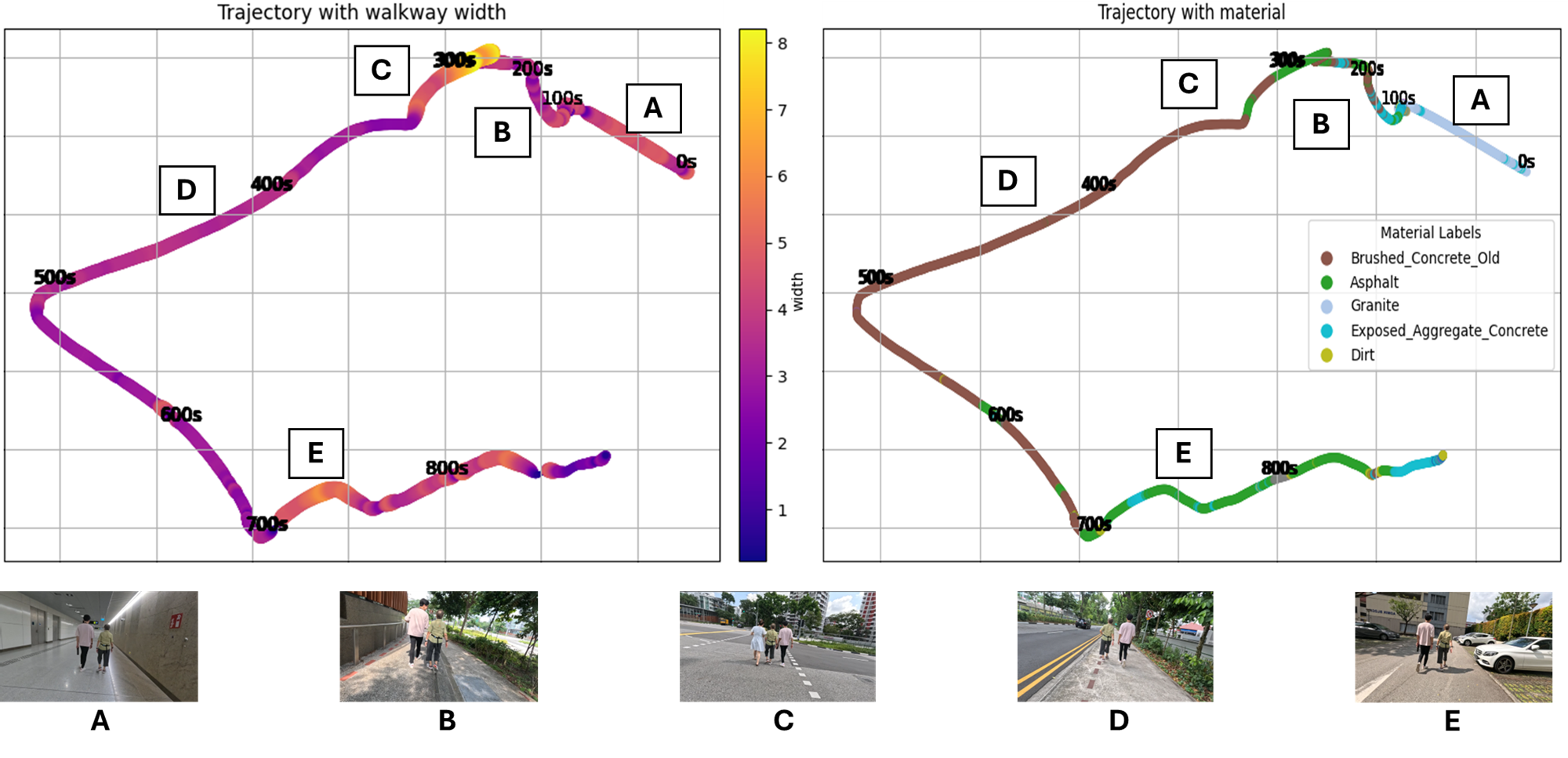} 
\caption{Reconstructed walking trajectories annotated with walkway attributes. (Left) Trajectory color-coded by estimated walkway width, showing transitions between narrow and wide path segments. (Right) Trajectory labeled with recognized walkway surface materials, including brushed concrete, asphalt, granite, and exposed aggregate concrete. Sample third-person video frames (bottom) illustrate representative walking contexts corresponding to points A–E along the trajectory.}
\label{fig3}
\end{figure*}

\section{Representation of Built Environments}
Accurate trajectory reconstruction is fundamental for analyzing human mobility and linking behavioral responses to specific spatial contexts. To obtain fine-grained participant trajectories, this study applies simultaneous localization and mapping (SLAM) using the VINS-Fusion algorithm on chest-mounted video recordings \cite{RN4}. Because GPS signals often degrade under dense canopies or near high-rise buildings, reliable GPS segments are selected as spatial anchors. Each SLAM-derived local trajectory is aligned with global geographic coordinates using the Umeyama similarity transformation, which combines the global accuracy of GPS with the local precision of SLAM \cite{umemaya}. The resulting continuous trajectory provides a accurate spatial foundation for mapping behavioral and BE features along each walking route.

Building on the reconstructed trajectory, multimodal data streams are analyzed to represent how participants perceive and interact with the BE. Gaze behavior offers a cognitive window into environmental perception. Fixations and saccades are detected using the I-DT algorithm with adaptive thresholds and optical-flow correction to compensate for head motion. Each fixation is projected onto corresponding scene frames and intersected with Mask2Former-segmented objects derived from the Mapillary Vistas dataset \cite{DBLP:journals/corr/abs-2112-01527}, allowing identification of which BE elements, such as signage, vegetation, or curbs, attract visual attention. Aggregated metrics, including fixation frequency, mean duration, and horizontal or vertical dispersion, characterize attentional engagement and scanning strategies.

Physiological responses further complement perceptual data. From the Empatica wristband, EDA signals are decomposed into phasic and tonic components, and peaks are detected to index arousal frequency. HRV metrics, such as Root Mean Square of Successive Differences (RMSSD) and pNN10, quantify autonomic regulation. These physiological indicators are temporally mapped onto the reconstructed trajectory to identify stress-responsive micro-environments, such as narrow corridors or uneven crossings that elicit heightened arousal.

Kinematic information derived from IMUs on the feet, wrist, and waist enables temporal stride segmentation and gait parameterization. Step count, mean stride time, and stride-time variability (STV) are computed to evaluate gait stability. Elevated STV or asymmetric step patterns may indicate adaptive responses to environmental obstacles or surface discomfort, particularly among participants with knee OA.

Among various BE features, walkways represent a critical element directly influencing older adults mobility. In this study, walkway surface materials are automatically classified using Grounding DINO \cite{RN69} for region proposals, Segment Anything Model (SAM) \cite{RN3} for precise mask extraction, and a linear probe for material classification across 14 urban categories (e.g., concrete, asphalt, tiles, bricks). Additionally, walkway width estimation leverages OpenPose \cite{RN6} detected skeletal keypoints: the pixel distance between walkway edges at the participant’s foot level is converted to meters using body-height references. These attributes quantify spatial and surface conditions that influence mobility confidence and gait adaptation.

Overall, all modalities are aligned to form time-based or distance-based feature segments. The fused multimodal dataset produces a spatiotemporal representation of experienced urban segments, enabling correlation analysis and visual mapping of how BE relate to behavioral, perceptual, and physiological responses.

\section{Preliminary Results}
Accurately capturing the dynamic interactions between older adults and complex BE requires a spatial mapping method capable of maintaining continuity across diverse conditions. This study addresses this challenge by leveraging a trajectory reconstruction and fusion approach that links global positioning with micro-scale tracking to represent mobility in real-world contexts.  The trajectory reconstruction was evaluated in representative urban settings in Singapore, including covered walkways, underground linkways, and semi-enclosed public spaces where GPS signals are frequently weak or unavailable.  As shown in Figure \ref{fig2}, SLAM provides smooth and locally consistent motion tracking even in GPS-denied environments, while GPS anchors the trajectory to global coordinates to preserve spatial accuracy. The fused trajectories exhibit continuous and stable motion profiles across indoor–outdoor transitions, accurately capturing direction changes and path curvature at sub-meter resolution. This ensures seamless spatial continuity and supports reliable mapping of behavioral and environmental metrics along the walking routes.

In addition, third-person video recordings captured from a chest-mounted GoPro were analyzed to extract walkway characteristics encountered during participants’ walking sessions. OpenPose was used to automatically track participants and detect foot positions, identifying regions of foot–ground contact in each frame. Walkway surfaces were segmented using Grounding DINO and the SAM, enabling precise localization of the areas directly walked on. These segmented patches formed the basis for two key analyses: walkway material classification and width estimation. Figure \ref{fig3} illustrate examples of material recognition and width estimation mapped onto the reconstructed trajectory. The visualized trajectories reveal clear transitions between narrow and wide walkway segments and varying surface materials, enabling spatiotemporal mapping of walkway conditions. These automated measurements provide fine-grained descriptors of walkway geometry and texture, establishing a foundation for analyzing how older adults adapt their gait and navigation strategies in response to different built-environment conditions.

\section{Discussion}
This study envisions a dynamic, multimodal approach to understanding how vulnerable older adults interact with the BE. A multimodal data fusion framework was developed that integrates wearable sensing (gaze, gait, physiological data) with environmental perception (video and GPS/SLAM mapping). This integration enables fine-grained quantification of human responses alongside environmental attributes such as walkway material, width, and visual openness. Preliminary results show that combining AI-based vision models with synchronized sensing streams allows reliable reconstruction of walking trajectories and detection of behavioral and physiological variations associated with specific BE conditions. The resulting multimodal representation moves beyond traditional GIS or street-view analyses by linking first-person and third-person perceptual, behavioral, and physiological data. For vulnerable older adults, this experience-based mapping dynamically reveals how micro-scale environmental variations affect movement stability and emotional comfort, providing urban planners and designers with new evidence-based indicators of accessibility and comfort. For example, frequent EDA peaks and wide fixation dispersion in narrow corridors may signal increased cognitive load or perceived risk, suggesting the need to redesign walkway geometry or surface materials. 

\begin{figure}
\centering
\includegraphics[width=0.9\columnwidth]{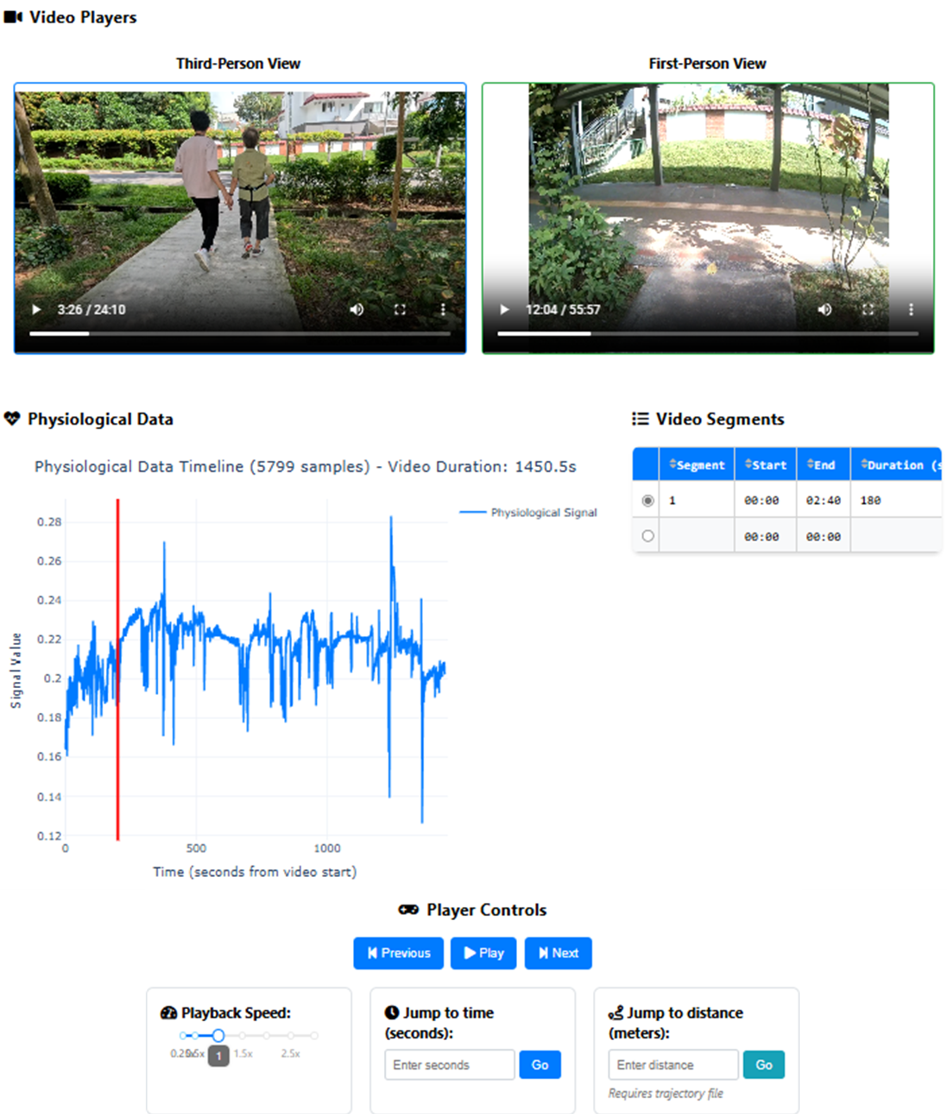} 
\caption{Prototype interactive dashboard for synchronized visualization of multimodal data streams. The interface displays third-person video, first-person video with fixation overlays, and physiological signals along the walking trajectory.}
\label{fig4}
\end{figure}

A prototype interactive dashboard visualizes synchronized data streams, including third-person video, first-person video with fixation overlays, and physiological signals, as shown in Figure \ref{fig4}. Users can navigate the data interactively by time or walking distance, enabling flexible exploration of the recorded sessions. This time–space visualization allows planners and researchers to identify locations where physiological arousal coincides with gaze fixations on specific BE elements, enabling intuitive interpretation of complex multimodal relationships between behavior, perception, and the BE. Furthermore, the fused representation can be integrated into urban digital twins to support holistic and aggregate evaluation of the BE. Visualizing trajectories from multiple participants enables hotspot analysis, helping uncover locations or design features that consistently elicit stress, attention shifts, or gait adaptation, thereby informing evidence-based urban design improvements.

\section{Conclusion and Future Work}
This study demonstrates the feasibility of multimodal data fusion to represent how BE features influence mobility, perception, and affective states in vulnerable older adults. By synchronizing gaze, gait, physiological, and environmental data collected in real urban settings, the approach is capable of generating dynamic and fine-grained representations of experienced urban spaces. These representations bridge human sensing and urban analytics, providing urban planners with quantitative insights into accessibility and comfort at the micro scale.

Future work will expand the participant pool to investigate statistical association and enable more comprehensive analyses of correlations across modalities, particularly between BE features and gait patterns in vulnerable populations. Integrating these multimodal outputs into city-scale digital twins will deepen the understanding of how the BE influences mobility among vulnerable groups and enable hotspot visualization and policy making. Overall, this human-centered and data-driven paradigm advances inclusive urban planning, contributing to the development of safer, more comfortable, and age-friendly environments for all.


\bibliography{workshop}

\end{document}